\documentclass{article}
\usepackage{spconf,amsmath,epsfig}

\newlength{\Lpr}
\newsavebox{\Bpr}

\newcommand{\D}[1]{\ensuremath{\displaystyle #1}}

\newcommand{\V}[1]{\mbox{\boldmath$\mathbf{#1}$\unboldmath}}

\newcommand{\sinc}{\ensuremath{{\mathrm{sinc}}}}

\newcommand{\bdm}{\begin{displaymath}}
\newcommand{\edm}{\end{displaymath}}

\newcommand{\be}[1]{\begin{equation} \label{#1}}
\newcommand{\ee}{\end{equation}}

\newcommand{\bae}[3]{
\begin{equation} \label{#1}
\renewcommand{\arraystretch}{#2}
\begin{array}{#3}}

\newcommand{\eae}{\end{array}\end{equation}}

\newcommand{\baen}[2]{
\begin{displaymath} 
\renewcommand{\arraystretch}{#1}
\begin{array}{#2}}

\newcommand{\eaen}{\end{array}\end{displaymath}}

\newcommand{\DefLetter}[4]{
\newcommand{#1}{\ensuremath{\V{#2}}} 
\newcommand{#3}{\ensuremath{\V{#4}}} 
}

\DefLetter{\vzer}{0}{\mzer}{0}
\DefLetter{\vone}{1}{\mone}{1}
\DefLetter{\va}{a}{\ma}{A}
\DefLetter{\vb}{b}{\mb}{B}
\DefLetter{\vc}{c}{\mc}{C}
\DefLetter{\vd}{d}{\md}{D}
\DefLetter{\ve}{e}{\me}{E}
\DefLetter{\vf}{f}{\mf}{F}
\DefLetter{\vg}{g}{\mg}{G}
\DefLetter{\vh}{h}{\mh}{H}
\DefLetter{\vi}{i}{\mi}{I}
\DefLetter{\vj}{j}{\mj}{J}
\DefLetter{\vk}{k}{\mk}{K}
\DefLetter{\vl}{l}{\ml}{L}
\DefLetter{\vm}{m}{\mm}{M}
\DefLetter{\vn}{n}{\mn}{N}
\DefLetter{\vpr}{p}{\mpr}{P}
\DefLetter{\vq}{q}{\mq}{Q}
\DefLetter{\vr}{r}{\mr}{R}
\DefLetter{\vs}{s}{\ms}{S}
\DefLetter{\vt}{t}{\mt}{T}
\DefLetter{\vur}{u}{\mur}{U}
\DefLetter{\vv}{v}{\mv}{V}
\DefLetter{\vw}{w}{\mw}{W}
\DefLetter{\vx}{x}{\mx}{X}
\DefLetter{\vy}{y}{\my}{Y}
\DefLetter{\vz}{z}{\mz}{Z}

\DefLetter{\vdel}{\delta}{\mdel}{\Delta}
\DefLetter{\vphi}{\phi}{\mphi}{\Phi}
\DefLetter{\vpsi}{\psi}{\mpsi}{\Psi}
\DefLetter{\vrho}{\rho}{\mrho}{\Lambda}
\DefLetter{\vxi}{\xi}{\mxi}{\Xi}

\DefLetter{\valpha}{\alpha}{\malpha}{\Alpha}
\DefLetter{\vbeta}{\beta}{\mbeta}{\Beta}
\DefLetter{\vlam}{\lambda}{\mlam}{\Lambda}
\DefLetter{\vsig}{\sigma}{\msig}{\Sigma}
\DefLetter{\vtau}{\tau}{\mtau}{\tau}
\DefLetter{\vtheta}{\theta}{\mtheta}{\Theta}
\DefLetter{\vome}{\omega}{\mome}{\Omega}
\DefLetter{\vzero}{0}{\mzero}{0}
\DefLetter{\vgam}{\gamma}{\mgam}{\Gamma}
\DefLetter{\veps}{\epsilon}{\meps}{\Epsilon}
\DefLetter{\veta}{\eta}{\meta}{\Eta}

\newcommand{\DefFuncLetter}[2]{
\newcommand{#1}{\ensuremath{{\mathrm{#2}}}} 
}

\DefFuncLetter{\Fzer}{0}
\DefFuncLetter{\Fa}{a}
\DefFuncLetter{\FA}{A}
\DefFuncLetter{\Fb}{b}
\DefFuncLetter{\Fc}{c}
\DefFuncLetter{\FC}{C}
\DefFuncLetter{\Fd}{d}
\DefFuncLetter{\Fe}{e}
\DefFuncLetter{\Ff}{f}
\DefFuncLetter{\Fg}{g}
\DefFuncLetter{\FG}{G}
\DefFuncLetter{\Fh}{h}
\DefFuncLetter{\FH}{H}
\DefFuncLetter{\Fi}{i}
\DefFuncLetter{\Fk}{k}
\DefFuncLetter{\Fl}{l}
\DefFuncLetter{\FL}{L}
\DefFuncLetter{\Fm}{m}
\DefFuncLetter{\Fn}{n}
\DefFuncLetter{\Fnr}{n}
\DefFuncLetter{\FN}{N}
\DefFuncLetter{\Fo}{o}
\DefFuncLetter{\FO}{O}
\DefFuncLetter{\Fpr}{p}
\DefFuncLetter{\FPr}{P}
\DefFuncLetter{\Fq}{q}
\DefFuncLetter{\Fr}{r}
\DefFuncLetter{\Fs}{s}
\DefFuncLetter{\FS}{S}
\DefFuncLetter{\Ft}{t}
\DefFuncLetter{\FT}{T}
\DefFuncLetter{\Fu}{u}
\DefFuncLetter{\FU}{U}
\DefFuncLetter{\Fv}{v}
\DefFuncLetter{\Fw}{w}
\DefFuncLetter{\FW}{W}
\DefFuncLetter{\Fx}{x}
\DefFuncLetter{\Fy}{y}
\DefFuncLetter{\FY}{Y}
\DefFuncLetter{\Fz}{z}
\DefFuncLetter{\FZ}{Z}
\DefFuncLetter{\Falp}{\alpha}
\DefFuncLetter{\Fbet}{\beta}
\DefFuncLetter{\Fchi}{\chi}
\DefFuncLetter{\Fdel}{\delta}
\DefFuncLetter{\Fzet}{\zeta}
\DefFuncLetter{\Feta}{\eta}
\DefFuncLetter{\Fphi}{\phi}
\DefFuncLetter{\FPhi}{\Phi}
\DefFuncLetter{\Fpsi}{\psi}
\DefFuncLetter{\FPsi}{\Psi}
\DefFuncLetter{\Fgam}{\gamma}
\DefFuncLetter{\FGam}{\Gamma}
\DefFuncLetter{\Flam}{\lambda}
\DefFuncLetter{\Fsig}{\sigma}
\DefFuncLetter{\Ftau}{\tau}
\DefFuncLetter{\Fome}{\omega}
\DefFuncLetter{\Feps}{\epsilon}
\DefFuncLetter{\Fthe}{\theta}
\DefFuncLetter{\Fvar}{\vartheta}

\DefFuncLetter{\FB}{B}
\DefFuncLetter{\FD}{D}
\DefFuncLetter{\FE}{E}
\DefFuncLetter{\FF}{F}
\DefFuncLetter{\FI}{I}
\DefFuncLetter{\FJ}{J}
\DefFuncLetter{\FM}{M}
\DefFuncLetter{\FR}{R}
\DefFuncLetter{\FX}{X}

\newcommand{\DefCalLetter}[2]{
\newcommand{#1}{\ensuremath{\mathcal{#2}}} 
}
\DefCalLetter{\CC}{C}
\DefCalLetter{\CD}{D}

\DefCalLetter{\CS}{S}
\DefCalLetter{\CV}{V}

\newcommand{\DefSubLetter}[2]{
\newcommand{#1}{\mathrm{#2}} 
}

\DefSubLetter{\slzer}{0}
\DefSubLetter{\sla}{a}
\DefSubLetter{\slA}{A}
\DefSubLetter{\slb}{b}
\DefSubLetter{\slB}{B}
\DefSubLetter{\slc}{c}
\DefSubLetter{\slC}{C}
\DefSubLetter{\sld}{d}
\DefSubLetter{\slD}{D}
\DefSubLetter{\sle}{e}
\DefSubLetter{\slE}{E}
\DefSubLetter{\slf}{f}
\DefSubLetter{\slF}{F}
\DefSubLetter{\slg}{g}
\DefSubLetter{\slG}{G}
\DefSubLetter{\slh}{h}
\DefSubLetter{\slH}{H}
\DefSubLetter{\sli}{i}
\DefSubLetter{\slI}{I}
\DefSubLetter{\slk}{k}
\DefSubLetter{\sll}{l}
\DefSubLetter{\slL}{L}
\DefSubLetter{\slm}{m}
\DefSubLetter{\slM}{M}
\DefSubLetter{\sln}{n}
\DefSubLetter{\slnr}{n}
\DefSubLetter{\slN}{N}
\DefSubLetter{\slo}{o}
\DefSubLetter{\slp}{p}
\DefSubLetter{\slP}{P}
\DefSubLetter{\slq}{q}
\DefSubLetter{\slQ}{Q}
\DefSubLetter{\slr}{r}
\DefSubLetter{\slR}{R}
\DefSubLetter{\sls}{s}
\DefSubLetter{\slS}{S}
\DefSubLetter{\slt}{t}
\DefSubLetter{\slT}{T}
\DefSubLetter{\slu}{u}
\DefSubLetter{\slU}{U}
\DefSubLetter{\slv}{v}
\DefSubLetter{\slw}{w}
\DefSubLetter{\slW}{W}
\DefSubLetter{\slx}{x}
\DefSubLetter{\slX}{X}
\DefSubLetter{\sly}{y}
\DefSubLetter{\slY}{Y}
\DefSubLetter{\slz}{z}
\DefSubLetter{\slZ}{Z}

\DefSubLetter{\slalp}{\alpha}
\DefSubLetter{\slbet}{\beta}
\DefSubLetter{\sldel}{\delta}
\DefSubLetter{\slDel}{\Delta}
\DefSubLetter{\sleps}{\epsilon}
\DefSubLetter{\slgam}{\gamma}
\DefSubLetter{\slphi}{\phi}
\DefSubLetter{\sltau}{\tau}
\DefSubLetter{\slxi}{\xi}
\DefSubLetter{\slthe}{\theta}

\title{Efficient Maximum Likelihood   Estimation of a 2-D    Complex Sinusoidal Based   on
  Barycentric Interpolation }
\name{J.  Selva\thanks{This   work   has been   accepted  for  publication   by  the  IEEE
    (International Conference on Acoustics, Speech and  Signal Processing 2011). Copyright
    may be   transferred  without notice,   after  which this  version  may  no  longer be
    accessible.  The author is with the Dept.  of Physics,  Systems Engineering and Signal
    Theory (DFISTS), University of Alicante, P.O.Box  99, E-03080 Alicante, Spain (e-mail:
    jesus.selva@ua.es).  This work has been supported by the Spanish Ministry of Education
    and Science (MEC), Generalitat Valenciana (GV), and by the University of Alicante (UA).
      }}

\address{Dpt. of Physics, Systems Engineering, and Signal Theory (DFISTS)\\
  University of Alicante,
  P.O. Box 99, \\
  03080 Alicante, Spain. E-mail: Jesus.Selva@ua.es }

\begin{document}
\maketitle
\ninept

\begin{abstract}

  This   paper presents an   efficient  method to  compute   the  maximum likelihood  (ML)
  estimation of the parameters of a complex  2-D sinusoidal, with  the complexity order of
  the  FFT.  The  method is  based  on an  accurate barycentric  formula for interpolating
  band-limited signals,  and on the  fact that the  ML cost  function  can be viewed  as a
  signal of this  type, if the  time  and frequency variables   are switched.  The  method
  consists in first computing the DFT of  the data samples, and  then locating the maximum
  of the cost function by means of Newton's algorithm.  The fact is that the complexity of
  the latter  step is small and  independent of the  data size, since it  makes use of the
  barycentric formula for obtaining the  values of the cost  function and its derivatives.
  Thus,  the total complexity  order is that of  the  FFT.  The method   is validated in a
  numerical example.
 
\end{abstract}

\keywords{Frequency estimation, parameter estimation, fast Fourier transform (FFT),
  barycentric interpolation, sampling.}

\section{Introduction}

The estimation of the parameters of a complex  sinusoidal in either  one or two dimensions
is a pervasive problem in signal processing. For this problem, the maximum likelihood (ML)
principle provides  an optimal estimator,  in the sense   that it achieves  the Cramer-Rao
bound at relatively low signal-to-noise ratios.  Yet in  practice this estimator is deemed
too complex, due to the associated maximization problem.  In rough terms, the situation is
that, though the ML cost function can be regularly sampled  using the FFT efficiently, the
localization of its  maximum requires some search  procedure, and here  is where the  high
computational burden seems unavoidable.  This observation was first stated  by B. G. Quinn
in \cite{Quinn91}, where it was shown that  a Gauss-Newton iteration may  fail to find the
global maximum of the cost function, if the initial iteration is taken from the DFT of the
data samples (without  zero padding).  This  has led several researches  to give up the ML
approach, and   look   for sub-optimal  estimators    with  lower  computational   burden,
\cite{Quinn94,So10}.  However,  in some references \cite{Aboutanios05, MacLeod98,Perisa06}
there has been an attempt to recover the initial ML approach, based on two arguments.  The
first is that it is feasible to approximately locate the maximum of  the ML cost function,
simply by looking for the DFT sample  with largest module.  Besides,  the accuracy of this
coarse  localization can be  improved  by zero padding.    And the second  is  that it  is
possible to interpolate  the cost  function  close to this  location  from the nearby  DFT
samples with some accuracy, since it is  a smooth function.   These arguments, if properly
exploited, make it possible  to improve the performance  significantly, with a  complexity
similar to that of the FFT.  The purpose of this paper is  to go one  step further in this
direction, and show that it is feasible  to perform this  interpolation with high accuracy
from a small number of  DFT samples, so  that the actual ML  estimate can be obtained with
the complexity of a single FFT.  The key lies in viewing the DFT  as a band-limited signal
in the frequency variable.

The basic concept in this paper is first presented in  the next section  in the context of
time-domain interpolation.  It is   a  simple an   efficient  technique to interpolate   a
band-limited  signal  from a  small number  of   samples with  high  accuracy.   Then, the
estimation problem  in one dimension  is introduced  in  Sec. \ref{sec:mfe1}, where  it is
shown that the method in the next section is  actually the key  for an efficient solution,
if the  independent  variable is properly  interpreted.   Afterward, Sec.   \ref{sec:mfe2}
presents the extension of the method  in Sec. \ref{sec:mfe1}  to the two-dimensional case,
and finally Sec.  \ref{sec:ne} contains a numerical example.

\section{High accuracy interpolation of a band-limited signal and its derivatives}

Given a  band-limited signal $\Fs(t)$ with  two-sided bandwidth  $B$,  a usual task  is to
interpolate its value from a finite set of samples surrounding $t$. If the sampling period
is $T$ with $BT<1$ (Nyquist  condition), and $2P+1$ samples  are taken symmetrically around
$t$, then this task consists in finding a set of coefficients $a_p(u)$ such that the formula
\be{eq:20}
\Fs(t)\approx\sum_{p=-P}^P \Fs((n-p)T)a_p(u)
\ee
is accurate, where  $t=nT+u$  is the  modulo-$T$  decomposition of $t$,   
\be{eq:50} n\equiv\lfloor t/T+1/2\rfloor,\;\; u\equiv t-nT.  \ee
For   fixed $u$,  the  $a_p(u)$  can be   obtained  numerically using  filter optimization
techniques \cite{Laakso96}. Yet this approach is cumbersome, if not only $\Fs(t)$ but also
its derivatives  must be interpolated for  varying values of  $u$ efficiently. Recently, a
so-called barycentric interpolator was derived in  \cite{Selva10} that solves this problem
satisfactorily. This interpolator takes the form
\be{eq:46}
\Fs(nT+u)\approx\left.\sum_{p=-P}^P \frac{\Fs((n-p)T)w_p}{u-pT}\right/\sum_{p=-P}^P 
\frac{w_p}{u-pT},
\ee
where $w_p$  is  a set  of  constants which  are samples  of  a fixed   function $\Fw(t)$,
[$w_p\equiv\Fw(pT)$].  In \cite{Selva10}, this function is given by the formula
\be{eq:22}
\Fw(t)\equiv \frac{\FGam(t/T+P+1)\FGam(-t/T+P+1)\Fg(t)}{\FL'(t)},
\ee
where $\FGam(\cdot)$ is the Gamma function, $\Fg(t)$ is the pulse 
\be{eq:23}
\frac{\sinc((1-BT) \sqrt{(t/T)^2-(P+1)^2)}}{\sinc(j (1-BT)(P+1))},
\ee
and $\FL(t)$ is the polynomial
\be{eq:24}
\FL(t)\equiv \prod_{p=-P}^P (t-pT).
\ee
(See \cite{Selva10}  for further details.)  The  fact is that  the error  of (\ref{eq:20})
decreases  exponentially with trend  $\Fe^{-\pi(1-BT)P}$. In practice  this  means that a
small $P$  is enough to  obtain high accuracy.  Besides,  as shown  in \cite{Selva10}, Eq.
(\ref{eq:20})   can  be differentiated with   low  complexity, so   as  to interpolate the
differentials of $\Fs(t)$ of any order, and (\ref{eq:46}) can  be evaluated using only one
division. This interpolator will be the fundamental tool in the next section.

\section{ML frequency estimation: 1-D case}
\label{sec:mfe1}

Consider  a signal $\Fz(t)$ consisting  of an undamped  exponential with complex amplitude
$a$ and frequency $f_o$, contaminated by complex  white noise $\Fn(t)$  of zero mean and
variance $\sigma^2$,
\be{eq:25}
\Fz(t)= \Fe^{j2\pi f_o t}+\Fn(t).
\ee
For simplicity, $f_o$ is assumed to lie in $[-1/2,1/2[$.  Next, assume that $M$ samples of
$\Fz(t)$ are  taken at instants  $m_1, m_1 +1,   \ldots, m_1+M-1$ with  $m_1\equiv -\lceil
M/2\rceil$. The maximum likelihood estimation of $f_o$ from these  samples is the argument
that maximizes the cost function
\be{eq:26}
\FL(f)\equiv |\Fc(f)|^2,
\ee
where $\Fc(f)$ is the correlation
\be{eq:27}
\Fc(f)\equiv \sum_{m=m_1}^{m_1+M-1} \Fz(m)\Fe^{-j2\pi f m}.
\ee
Since $\Fc(f)$   is the DFT   of the samples  $\Fz(m)$,  the maximum  of  $\FL(f)$ can be
approximately  located by selecting a  frequency spacing $\Delta  f$  such that $1/\Delta
f$ is a power of two, and then computing the samples
\be{eq:28}
\Fc(k\Delta f)\equiv \sum_{m=m_1}^{m_1+M-1} \Fz(m)\Fe^{-j2\pi k m \Delta f },
\ee
by means of a radix-2 FFT algorithm.  The cost  of this operation  is only $\FO(M\log M)$.
This way, it is possible to obtain a frequency $k_o\Delta  f$ that lies  close to the true
abscissa  of the maximum  of  $\FL(f)$.  However,  if  $\hat{f}_o$ is  this abscissa,  the
approximation of  $\hat{f}_o$ using $k_o\Delta  f$ is very  inaccurate. In this situation,
the  accuracy can  be improved  by   reducing $\Delta  f$   (over-sampling), but then  the
computational burden becomes high.

The  barycentric  interpolator in  the previous  section  provides an  efficient method to
obtain $\hat{f}_o$ The key  idea is that  the correlation $\Fc(f)$  in (\ref{eq:27}) is  a
band-limited signal \emph{in the $f$ variable} of bandwidth  $2m_1$, i.e, the variable $f$
in (\ref{eq:27}) plays the same role as the variable $t$ in (\ref{eq:20}).  Also, $\FL(f)$
is band-limited  with bandwidth $4m_1$. So,  the method of  sampling the correlation using
the FFT, and then looking for its maximum  module can be  reformulated taking into account
these information.  First, since $\FL(f)$ has two-sided  bandwidth $4m_1$, it is necessary
to sample this function with a  spacing smaller than  its Nyquist period  $1/(4 m_1 )$, in
order to  coarsely locate its  maximum.  This implies  that  a factor-two  zero padding is
enough to ensure the localization of the maximum of $\FL(f)$.   And second, since $\Fc(f)$
has   bandwidth   $2m_1$, it  can   be  interpolated  using  the   barycentric  formula in
(\ref{eq:20}) with $2m_1$ in place of $B$, and $\Delta f$ in place of $T$, i.e,
\be{eq:29}
\Fc(f)
\approx\left.\sum_{p=-P}^P \frac{\Fc((k-p)\Delta f)w_p}{\gamma-p\Delta f}\right/
\sum_{p=-P}^P 
\frac{w_p}{\gamma-p\Delta f},
\ee
where $f=k\Delta  f+\gamma$ is the  modulo-$\Delta f$ decomposition of  $f$  defined as in
(\ref{eq:50}). If $\tilde{\Fc}(f)$ denotes the approximation to $\Fc(f)$ in (\ref{eq:29}),
then one may replace the problem of maximizing $\tilde{\FL}(f)$ with that of maximizing
\be{eq:34}
\tilde{\FL}(f)\equiv |\tilde{\Fc}(f)|^2
\ee
Besides, the maximum of this function is close to  the abscissa of  the largest FFT sample
in (\ref{eq:28}), $k_o\Delta f$.

Since  $P$  is small  and  there  is  a  coarse  estimate  of  the  maximum  abscissa, the
maximization  of (\ref{eq:34}) is  a  low-complexity  problem  that  can  be solved  using
standard numerical methods.  Given that the differentials  of the barycentric interpolator
can be  easily computed \cite[Sec.  IV]{Selva10}, a suitable  numerical method is Newton's
algorithm, in which the $r$th iteration $f_r$ is refined using
\be{eq:35}
f_{r+1}=f_r+\tilde{\FL}'(f_r)/\tilde{\FL}''(f_r)
\ee
where 
\bae{eq:36}{2}{r@{\,=\,}l}
\D{\tilde{\FL}'(f)}&\D{\frac{2}{M}\mathrm{Re}\{\tilde{\Fc}'(f)\tilde{\Fc}(f)^*\},}\\ 
\D{\tilde{\FL}''(f)}&\D{\frac{2}{M}(\mathrm{Re}\{\tilde{\Fc}''(f)\tilde{\Fc}(f)^*\}
+|\tilde{\Fc}'(f)|^2).}
\eae
This process can be initiated with  $f_1=k_o\Delta f$, and a  small number of iterations
(3 to 5) is enough to obtain $\hat{f}_o$. 

The computational burden of this method is given by that of the FFT, i.e, it is $\FO(M\log
M)$, and it yields the actual ML estimate of $f_o$. 

\section{ML frequency estimation: 2-D case}
\label{sec:mfe2}

The method  in  the previous  section can  be  extended  to  two-dimensional  signals with
non-essential changes. For this, it is only necessary to consider two variables, $t_1$ and
$t_2$, and  repeat the  same  interpolation procedure.  The  initial model,  equivalent to
(\ref{eq:25}), is
\be{eq:37}
\Fz(t_1,t_2)=\Fe^{j2\pi (f_{1,o} t_1+f_{2,o}t_2)}+\Fn(t_1,t_2).
\ee
Next,  this signal is   sampled at the   integer pairs $(m,n)$   for $m_1\leq m<m_1+M$ and
$n_1\leq  n<n_1+N$, with $m_1\equiv-\lceil   M/2\rceil$, $n_1\equiv-\lceil N/2\rceil$. The
2-D equivalent of the cost function in (\ref{eq:26}) is
\be{eq:38}
\FL(f_1,f_2)\equiv |\Fc(f_1,f_2)|^2,
\ee
where $\Fc(f_1,f_2)$ is the correlation
\be{eq:39}
\Fc(f_1,f_2)\equiv \sum_{m=m_1}^{m_1+M-1} \sum_{n=n_1}^{n_1+N-1}\Fz(m,n)
\Fe^{-j2\pi (f_1 m+f_2n)}.
\ee
This function can be sampled with spacings $\Delta f_1$, $\Delta  f_2$ using a radix-2 2-D
FFT. These samples provide a  frequency pair $(k_{1,o}  \Delta f_1,$ $k_{2,o} \Delta f_2)$
that  lies close to  the maximum of $\FL(f_1,f_2)$.   Then, it  is possible  to  set up an
interpolation formula like (\ref{eq:29}) but in two dimensions,
\bae{eq:40}{1.5}{l}
\D{\Fc(f_1,f_2)\approx\sum_{p_1=-P_1}^{P_1} \sum_{p_2=-P_2}^{P_2} }\\
\D{{}\hspace{0.5cm}
\frac{\Fc((k_1-p_1)\Delta f_1,(k_2-p_2)\Delta f_2)w_{1,p_1}w_{2,p_2}}
{(\gamma_1-p_1\Delta f_1)(\gamma_1-p_1\Delta f_1)}\cdot}\\
\D{{}\hspace{0.9cm}
\Big[\sum_{p_1=-P_1}^{P_1} \frac{w_{1,p_1}}
{\gamma_1-p_1\Delta f_1}
\sum_{p_2=-P_2}^{P_2} 
\frac{w_{2,p_2}}{\gamma_2-p_2\Delta f_2}\Big]^{-1},
}
\eae
where $f_1=k_1\Delta   f_1+\gamma_1$ and  $f_2=k_2\Delta   f_2+\gamma_2$ are   the  modulo
decompositions, and $w_{1,p_1}$ and $w_{2,p_2}$ are  the barycentric weights corresponding
to bandwidths $2m_1$ and $2n_1$, and truncation indices $P_1$ and $P_2$, respectively.  If
$\tilde{\Fc}(f_1,f_2)$ denotes the  formula in  (\ref{eq:40}),  the problem of  maximizing
$\FL(f_1,f_2)$ can be substituted by the problem of maximizing
\be{eq:41}
\tilde{\FL}(f_1,f_2)\equiv |\tilde{\Fc}(f_1,f_2)|^2.
\ee
Finally, the Newton iteration for this problem is 
\be{eq:42}
(f_{1,r+1},f_{2,r+1})=(f_{1,r},f_{2,r})+\mathcal{H}\tilde{\FL}^{-1}\nabla
\tilde{\FL},
\ee
where $\nabla\tilde{\FL}$ and  $\mathcal{H}\tilde{\FL}$ are the  gradient and Hessian  of
$\tilde{\FL}$ respectively, evaluated at $(f_{1,r},f_{2,r})$. These functionals are
\be{eq:43}
\nabla\tilde{\FL}=\frac{2}{MN}\,\mathrm{Re}\{ \nabla\tilde{\Fc}\tilde{\Fc}^*\}
\ee
and
\be{eq:44}
\mathcal{H} \tilde{\FL}=\frac{2}{MN}\,\mathrm{Re}\{\mathcal{H}\tilde{\Fc}\,\tilde{\Fc}^*+
 \nabla \tilde{\Fc}\nabla\tilde{\Fc}^\mathrm{H}\},
\ee
where  $\nabla{\tilde{\Fc}}$ and $\mathcal{H}\tilde{\Fc}$  are  the corresponding gradient
and Hessian of $\tilde{\Fc}$.

Since the evaluation  of $\tilde{\Fc}(f_1,f_2)$ has  a small cost  which is independent of
either $M$ or $N$, the complexity is given  by the 2-D FFT,  i.e, it is $\FO(MN\log(MN))$.
Note that sub-optimal  methods like that  in \cite{So10} have complexity $\FO(MN(M+N))$,
that  is, the method  in this section yields  the ML estimate  and, besides, has a smaller
complexity order.

\section{Numerical examples}
\label{sec:ne}

Let $\Fphi(f,t)$ denote the  value delivered by the  barycentric formula in (\ref{eq:46}),
when the input signal  is  the undamped exponential   $\Fs(t)=\Fe^{j2\pi ft}$.  Since  the
functions to interpolate in this  paper are sums of  exponentials like $\Fe^{j2\pi ft}$, a
simple way to  assess the  interpolation  accuracy is to   evaluate the maximum  module of
$\Fe^{j2\pi fu}-\Fphi(f,u)$,  for $|u|\leq T/2$  and $f$ varying  in $[-B/2,B/2]$, i.e, to
assess the spectrum function
\be{eq:47}
\FE(f)\equiv \max_{|u|\leq T/2}|\Fe^{j2\pi f u}-\Fphi(f,u)|.
\ee
\begin{figure}
\includegraphics{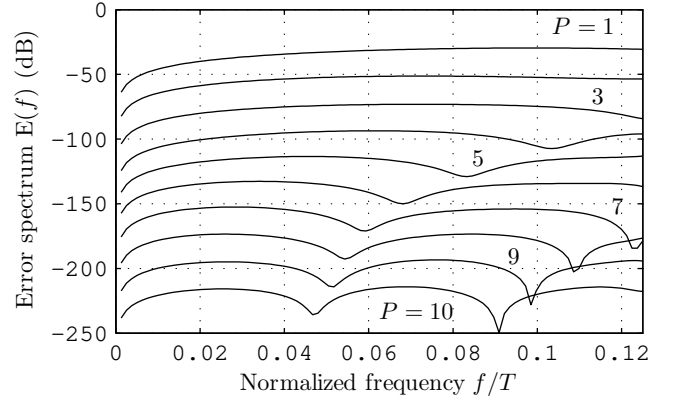}
\caption{\label{fig:4} Error spectrum $\FE(f)$ of the barycentric interpolator in
  (\ref{eq:46}) for several truncation indices $P$.}
\end{figure}
Fig \ref{fig:4} shows $\FE(f)$ for $BT=0.25$ and several truncation indices $P$. Note that
any accuracy can be achieved uniformly in $[-B/2,B/2]$, by slightly increasing $P$.

\begin{figure}
\includegraphics{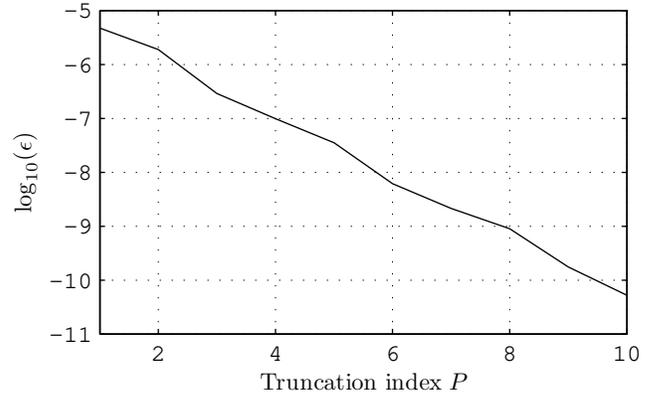}
\caption{\label{fig:2}  Truncation  index  $P$  versus the    $\log_{10}$ of  the  maximum
  interpolation error ($\log_{10}\epsilon$) for the cost function $\FL(f_1,f_2)$.}
\end{figure}

In order to test the interpolation error in a specific example, a 2-D undamped exponential
with $M=500$,   $N=651$ and  frequencies  $f_{o,1}=0.234452$  and  $f_{o,2}=-0.143254$ was
generated. Then,  complex white  noise  was added,  so  that the signal-to-noise  ratio was
$S/N=5$  dB.   Fig. \ref{fig:2}  shows   the interpolation  error for the cost function
$\FL(f_1,f_2)$,  defined by
\be{eq:45}
\epsilon\equiv \frac{\max_{f_1,f_2} |\FL(f_1,f_2)-\tilde{\FL}(f_1,f_2)|}
{\max_{f_1,f_2} \FL(f_1,f_2)},
\ee
for varying truncation index $P$. Again, it is clear that any accuracy can be achieved by
slightly increasing $P$. 

Next, two estimators were compared in this example. To describe them, let $\mz$ denote the
data matrix obtained by sampling (\ref{eq:37}) as described in that section,
\be{eq:48}
[\mz]_{m,n}\equiv \Fz((m_1+m-1)T_1,(n_1+n-1)T_2),
\ee
and let $\vur_1$, $\vv_1$ denote its first left and right singular vectors, respectively.
The first method consisted in computing the 1-D ML estimator from $\vur_1$ and $\vv_1$ so
as to obtain the respective estimations of $f_{o,1}$ and $f_{o,2}$,
\bae{eq:49}{1.5}{l}
\D{\hat{f}_{o,1}=\arg \max_f \Big|\sum_{m=1}^{M}[\vur_1]_m\Fe^{-j2\pi(m_1-1+m)f}\Big|^2.}\\
\D{\hat{f}_{o,2}=\arg \max_f \Big|\sum_{n=1}^{N}[\vv_1]_n\Fe^{-j2\pi(n_1-1+n)f}\Big|^2.}
\eae
The actual estimations $\hat{f}_{o,1}$ and $\hat{f}_{o,2}$ were computed from $\vur_1$ and
$\vv_1$ by applying the  1-D method in  Sec.  \ref{sec:mfe1} to  each of them. This method
had complexity $\FO(MN(M+N))$, since  it involved the  computation of the singular vectors
$\vur_1$ and $\vv_1$, and  is equivalent to  the method in  \cite{So10}. The second method
computed the actual ML estimator from  $\mz$ using the  method in Sec. \ref{sec:mfe2}, and
its complexity was just $\FO(MN\log(MN))$.  For the values of $M$ and $N$ given above, the
over-sampling factors were $2.048$ and  $1.523$, respectively, i.e, the  FFT had size 1024
in both dimensions.

\begin{figure}
\includegraphics{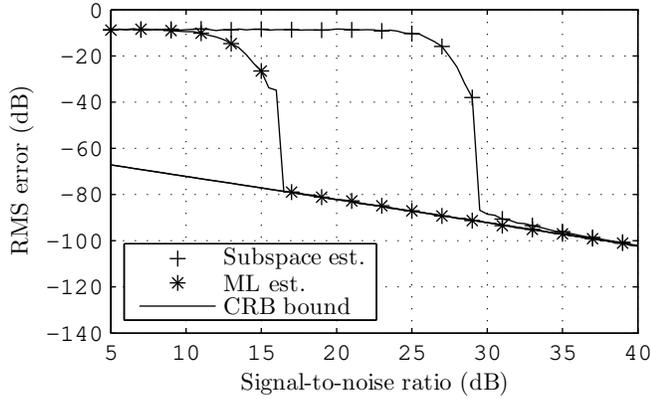}
\caption{\label{fig:1} Root-Mean-Square (RMS) error of the subspace and ML estimators,
  together with the Cramer-Rao bound.}
\end{figure}

Fig.  \ref{fig:1} shows the  root-mean-square (RMS) error  for the first estimator, termed
subspace estimator,   for  the second  estimator   (interpolated ML  estimator),   and the
Cramer-Rao bound. The threshold performance of the second is clearly superior.

\begin{figure}
\includegraphics{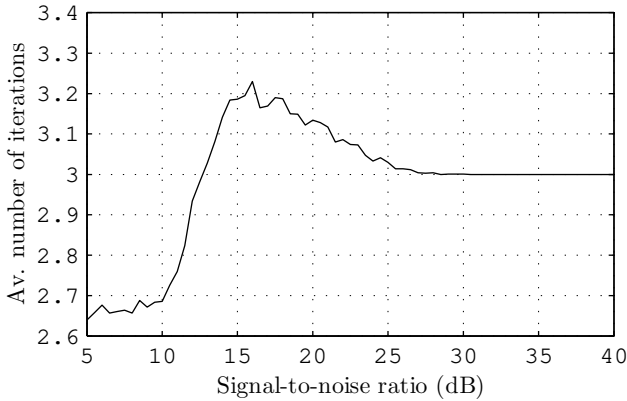}
\caption{\label{fig:3} Average number of iterations required by the Newton method.}
\end{figure}

Finally,  Fig.  \ref{fig:3} shows  the  average number  of   iterations that  required the
interpolated ML estimator, which was roughly equal to three. 

\section{Conclusions}

A  method  has been  presented that  allows  one to compute   the  maximum likelihood (ML)
estimation of a complex 2-D sinusoidal, with the complexity of  the fast Fourier transform
(FFT). First, it is recalled in the paper  that a band-limited  signal can be interpolated
with high accuracy from a small number  of samples, if  the sampling frequency is somewhat
higher than the Nyquist frequency. Besides, a specific barycentric  formula is proposed to
perform this kind of interpolation.  And second, it is shown that the ML cost function for
the  estimation of a  complex 2-D (and  1-D)  sinusoidal can be  viewed  as a band-limited
signal, if the time and frequency variables are switched.  Finally,  these two results are
combined in a method that is able to deliver the ML estimate  with the complexity order of
the FFT, which is based on Newton's algorithm.

\bibliographystyle{IEEEbib}

\bibliography{c:/JesusSelva/Data2/Utilities/LaTeX/Bibliography}

\end{document}